# Structural Complexity of One-Factor Sparse Portfolio Selection

*Exact Algorithms, Parameterized Hardness, and Restricted Circuit Lower Bounds*

**Davit Gondauri**[1,2]

[1] Full Professor, Doctor of Business Administration, Business and Technology University (BTU), Tbilisi, Georgia

[2] Director, Eurasian Logistics Research Center, European Marketing and Management Association (EUMMAS)

ORCID: 0000-0002-9611-3688

Correspondence: dgondauri@gmail.com

## Abstract

We study exact-cardinality, equally weighted minimum-variance portfolio selection under a one-factor covariance model supplied in factor form. In the nonnegative homoskedastic regime, selecting the K smallest loadings is optimal. Allowing strictly positive asset-specific idiosyncratic variances makes the decision problem NP-complete even with positive integer loadings and a strictly positive-definite covariance matrix; with identity residual covariance, exactly one negative loading also suffices. We give exact pseudo-polynomial dynamic programs for one factor and fixed factor dimension and prove W[1]-hardness parameterized by K, including the positive-data family. Consequently, a general exact polynomial-time algorithm for Monge's (2017) equally weighted single-factor variance-input formulation would imply P=NP. For a normalized binary factor encoding, we construct a depth-zero projection from modular k-SUM that preserves exact cardinality and positive definiteness. The projection transfers Lin's (2026) fixed-k circuit lower bound under its stated width and quantifier conditions and, independently, yields a parity-based proof that the portfolio language is not in nonuniform $AC^0$ even with identity residual covariance and polynomially bounded integer coefficients. These are restricted-circuit results: no unrestricted P/poly lower bound and no separation of P from NP is claimed.

## 1. Introduction, Decision Language, and Canonical Factor Encoding

Cardinality-constrained quadratic portfolio optimization has long been studied through mixed-integer and structured quadratic optimization methods (Bienstock, 1996; Bertsimas & Shioda, 2009). Low-rank quadratic 0–1 optimization also has established pseudo-polynomial and polynomially solvable regimes (Çela, Klinz, & Meyer, 2006; Gao & Li, 2013). Factor-model formulations remain active in portfolio research, including the equally weighted single-factor formulation of Monge (2017, 2020) and more recent decision-based factor-model approaches (Anis & Kwon, 2025). Ebrahimi, Amini, and Liu (2026) provide a recent NP-hardness result for a broader cardinality-constrained portfolio model. The present paper studies a narrower exact-cardinality, equally weighted, one-factor language and asks how its tractable regimes, NP-complete restrictions, parameterized complexity, and restricted-circuit complexity fit together.

Let n be the number of assets and $S \subseteq \{1,\ldots,n\}$ with $|S|=K$. Equal weighting gives $w_i=1/K$ for $i\in S$ and $w_i=0$ otherwise. The covariance is represented by its one-factor data, not by an explicit n×n matrix:

$$\Sigma = \beta\beta^{T} + D \qquad (1)$$

where $\beta=(\beta_1,\ldots,\beta_n)^T$ and $D=\mathrm{diag}(d_1,\ldots,d_n)$. Since multiplication of portfolio variance by $K^2$ does not change the optimizer, the scaled objective is

$$F(S) = (\textstyle\sum_{i\in S}\beta_i)^2 + \sum_{i\in S} d_i \qquad (2)$$

The factor-encoded decision language $L_1F$ takes integer data $(\beta,D,C)$ unless a later section explicitly states otherwise, and asks whether there exists S of exactly K indices with $F(S)\le C$. All hardness and circuit-transfer statements below concern this factor representation. An explicit-covariance-input variant, in which every entry of Σ is encoded directly, is a different language; the depth-zero projection of Section 8 does not automatically transfer to that variant because forming $\beta_i\beta_j$ would require arithmetic rather than a bit projection.

### 1.1 Unique normalized canonical encoding

Define nat(x) for $x\ge 0$ as follows. Let $\ell=1$ for $x=0$ and $\ell=\lfloor \log_2 x\rfloor +1$ for $x\ge 1$; encode x as 1 repeated ℓ times 0 followed by its ℓ-bit unsigned binary representation. This is the only prefix code used in the header. Define $\mathrm{swidth}(z)=\min\{w\ge 1: -2^{w-1}\le z\le 2^{w-1}-1\}$ and $\mathrm{uwidth}(y)=\min\{w\ge 1: 0\le y\le 2^{w}-1\}$.

For a numerical instance set $W\beta=\max_i \mathrm{swidth}(\beta_i)$, $Wd=\max_i \mathrm{uwidth}(d_i)$, and $WC=\mathrm{swidth}(C)$. The unique well-formed encoding is nat(n)||nat(K)||nat(Wβ)||nat(Wd)||nat(WC), followed by exactly n Wβ-bit two's-complement β-blocks, exactly n Wd-bit unsigned d-blocks, and one WC-bit two's-complement C-block. The declared widths must equal these minimal normalized widths. Any extra sign-extension or zero padding, any malformed prefix field, any mismatch in block count, or any instance with $n<1$ or $K\notin\{1,\ldots,n\}$ or $d_i<0$ is rejected. Therefore every numerical instance has exactly one canonical bit string and $L_1F$ is a total Boolean language on $\{0,1\}^*$.

For positive-definite subfamilies require $d_i>0$. Rational data arise only in the fixed-r extension. Encode a reduced rational p/q with q>0 by a sign bit followed by nat(|p|) and nat(q); zero has sign 0 and denominator 1. The one-factor hardness and circuit-transfer constructions use integer data only.

Encoding validation is polynomial in the supplied string length. Reject zero widths, inconsistent headers and impossible block counts before scanning data blocks, then verify their minimal widths. A certificate lists K distinct indices; exact integer arithmetic verifies the objective in polynomial time. Hence the total language, including its malformed-input convention, belongs to NP. Throughout, $\mathbb{Z}_+$ denotes strictly positive integers and logarithms in bit bounds have base two.

### 1.2 Positive definiteness

$$x^{T}\Sigma x = (\beta^{T}x)^2 + \textstyle\sum_i d_i\, x_i^2 > 0 \ (x\neq 0, d_i>0) \qquad (3)$$

Hence every construction below with $d_i>0$ lies inside the strictly positive-definite covariance class.

### 1.3 Contributions and significance

The paper develops a unified complexity map for one-factor sparse portfolio selection across specified structural regimes. Within one canonical factor-encoded language, it identifies a polynomial-time regime, sharply restricted NP-complete regimes, an exact pseudo-polynomial frontier, W[1]-hardness in the natural cardinality parameter, and an unconditional lower bound for nonuniform $AC^0$.

The circuit contribution is an explicit normalized modular k-SUM projection whose output coordinates copy source bits or constants. It supports two consequences: a fixed-k quantitative transfer from Lin (2026), and an independent full-language $AC^0$ separation from the classical parity lower bound of Furst, Saxe, and Sipser (1984). The latter already holds on polynomially bounded numerical instances; the quantitative fixed-k statement concerns a different parameter regime.

Two structural comparisons organize the decision results. Allowing asset-specific positive residual variances yields an NP-complete family with positive loadings, whereas constant residual variances and nonnegative loadings are tractable. Under D=I, permitting exactly one negative loading also yields an NP-complete family. These are worst-case family statements, not claims that every departure from homoskedasticity or nonnegativity creates a hard instance. Theorem 4 preserves K=k in a reduction to the positive-data family.

The novelty is structural and encoding-specific rather than generic. The manuscript does not claim the first rank-one quadratic NP-hardness proof, a new parity lower bound, NP⊄P/poly, or P≠NP. Its substantive contributions are the portfolio-specific restrictions in Theorems 2 and 6, the parameter-preserving positive-data reduction in Theorem 4, and the normalized bit projection in Theorem 7. Section 9 records only the standard P/poly consequence of NP-completeness and uses it to delimit the unrestricted circuit frontier.

## 2. Tractable Nonnegative Homoskedastic Regime

**Theorem 1 (nonnegative homoskedastic tractability).** If $d_i=d>0$ and $\beta_i\geq 0$ for all i, an optimal exactly-K equally weighted subset consists of the K smallest $\beta_i$ values. Sorting requires O(n log n) comparisons. For binary input of total length L, this is a polynomial-time algorithm; its bit cost includes reading the input and comparing multi-bit integers, and is not asserted to be O(n log n) bit operations.

Proof. For every feasible S, $F(S)=(\sum_{i\in S}\beta_i)^2+Kd$. The second term is constant, and the square is increasing on $[0,\infty)$. Thus minimizing F is equivalent to minimizing the nonnegative loading sum, achieved by the K smallest loadings. □

## 3. Positive One-Factor NP-Completeness with Asset-Specific Idiosyncratic Risk

**Lemma 2.1 (positive Exact-K Subset Sum is NP-complete).** Given $n\geq 1$ positive integers $a_1,\ldots,a_n$, a positive integer target t, and $1\leq K\leq n$, deciding whether exactly K distinct indices have values summing to t is NP-complete. Repeated values at different indices are permitted.

Proof. Reduce from SUBSET SUM on positive integers with positive target (Garey & Johnson, 1979). If t exceeds the sum of all source integers, output the fixed no-instance ($a_1'=1,t'=2,K'=1$). Otherwise create n integers $a_i+1$ and n additional integers equal to 1. Set $K'=n$ and $T=t+n$. Any source solution with j selected indices extends to exactly n target indices by adding $n-j$ of the unit items, and has target sum T. Conversely, every n-item target subset has sum n plus the sum of those $a_i$ whose shifted items were selected, and therefore yields a source solution precisely when its sum is T. All integers have polynomial binary length. Membership in NP follows by summing the chosen values. □

**Theorem 2 (positive-integer, strictly positive-definite one-factor NP-completeness).** $L_1F$ is NP-complete even when $\beta_i,d_i\in\mathbb{Z}_+$ for every i and $\Sigma=\beta\beta^T+D$ is positive definite.

Proof. Reduce from Lemma 2.1. Given $a_1,\ldots,a_n$,t and cardinality k, set K=k, $\beta_i=a_i$, $a_{max}=\max_i a_i$, $M=2ta_{max}+t^2+1$, $d_i=M-2ta_i$, and threshold $C=kM-t^2$. Then $d_i\geq t^2+1>0$. For a feasible S let $s=\sum_{i\in S}a_i$. We have $F(S)=s^2+kM-2ts=(s-t)^2+kM-t^2$. Therefore $F(S)\leq C$ iff s=t. The target has n items and all constructed integers have polynomial bit length. Membership in NP follows from direct polynomial-bit evaluation of F. □

$$F(S) = (s-t)^2 + kM - t^2 \qquad (4)$$

### 3.1 Structural classification and the 2017 EWCCMVSF question

**Corollary 2.1 (positive-loading structural tractability-hardness boundary).** Within the factor-encoded equally weighted one-factor family with nonnegative loadings, the homoskedastic regime $d_i=d>0$ is solvable in O(n log n) comparisons by Theorem 1, whereas allowing strictly positive idiosyncratic variances to vary across assets makes the family NP-complete by Theorem 2, even while $\beta_i>0$ and $\Sigma\succ 0$. This is a family-level structural boundary: it does not claim that every instance with unequal idiosyncratic variances is hard.

Model containment. Let $x_i=1$ precisely on S, set Monge's factor variance to one and residual variance $\sigma^2\varepsilon_i=d_i$. The equal weights $w_i=x_i/K$ give variance $V(S)=F(S)/K^2$. Thus $F(S)\leq C$ iff $V(S)\leq C/K^2$, with exactly K selected assets. All coefficients supplied by Theorem 2 are strictly positive and have polynomial binary length. Taking residual variances as input avoids any need to encode irrational square roots. This is a direct containment in the EWCCMVSF formulation, not a reduction from unrestricted covariance data.

**Corollary 2.2 (exact optimization consequence for EWCCMVSF).** An exact polynomial-time algorithm for the general EWCCMVSF optimization problem, on rational or integer factor data, would imply P=NP. Proof. Apply the algorithm to the positive-integer instances of Theorem 2. Either evaluate its returned subset exactly or compare its exact optimum with $C/K^2$. The comparison decides the NP-complete source problem in polynomial time. Therefore no such general exact polynomial-time algorithm exists unless P=NP. □

Monge (2017, Appendix, p. 22) asks whether EWCCMVSF is polynomially solvable. Corollary 2.2 answers this question for the general binary-encoded variance-input formulation: a polynomial-time exact algorithm would imply P=NP. The related 2020 journal article is cited for the factor-model context; the question addressed here is the one stated in the 2017 preprint.

## 4. Exact Dynamic Programming and Numerical Complexity

**Theorem 3 (exact pseudo-polynomial algorithm).** Assume $\beta_i,d_i\in\mathbb{Z}_+$ and let B be the sum of the loadings. The optimum value is computable with O(nKB) transitions and O(KB) stored numerical entries. For total binary input length L, the bit cost is O(nKB·poly(L)); O(KB) counts entries, not bits.

Define $DP_j(k,b)$ as the minimum attainable $\sum d_i$ among subsets of {1,…,j} with cardinality k and loading sum b, using $+\infty$ if none exists. The recurrence is

Initialization and boundary conventions. $DP_0(0,0)=0$; all other initial states are $+\infty$. An out-of-range cardinality or loading sum is assigned $+\infty$, and $+\infty+d=+\infty$. Each transition reads only the preceding j-layer, preventing reuse of the same asset. To output a witnessing subset as well as its value, retain predecessor information using additional memory or reconstruct by recomputation.

$$DP_j(k,b) = min\{DP_{j-1}(k,b), DP_{j-1}(k-1,b-\beta_j)+d_j\} \qquad (5)$$

and the exact optimum is

$$OPT = min_{0\leq b\leq B}\{b^2 + DP_n(K,b)\} \qquad (6)$$

Rolling arrays remove the j-dimension. Because B is a magnitude rather than its binary length, the algorithm is pseudo-polynomial.

**Corollary 3.1 (weak numerical NP-completeness).** The positive-integer family is NP-complete and has a pseudo-polynomial exact algorithm. In the standard numerical sense it is weakly NP-complete; unless P=NP, the same integer family is not strongly NP-hard under polynomially bounded magnitudes. No corresponding claim is made for unrestricted rational encodings, continuous weights, or additional constraints.

## 5. Parameterized W[1]-Hardness in K

Source theorem. Abboud, Lewi, and Williams (2014; arXiv:1311.3054) prove W[1]-completeness for distinct-index zero-sum k-SUM with integers in $[-n^{2k},n^{2k}]$. We use this zero-target formulation directly; no randomized reduction from unrestricted integers is needed. Source numbers have O(k log n) bits.

**Theorem 4 (parameter-preserving W[1]-hardness).** $L_1F$ is W[1]-hard when parameterized by K.

Proof. Given a bounded-number k-SUM instance $x_1,\ldots,x_n$ with target 0, let $R=\max_i|x_i|$, set $A=R+1$, $a_i=x_i+A>0$ and $t=kA$. Then every k-set T satisfies $\sum_{i\in T}a_i=t$ iff $\sum_{i\in T}x_i=0$. The transformation preserves the parameter exactly: K=k. Because $|x_i|\le n^{2k}$, all shifted integers use O(k log n) bits. Applying Theorem 2 yields an fpt many-one reduction with output parameter K=k and output bit length f(k)·poly(n). Thus $L_1F$ is W[1]-hard. □

Parameterized interpretation. Theorem 4 applies already to positive integer loadings and positive asset-specific residual variances. Assuming FPT≠W[1], it rules out running time f(K) times the c-th power of L for a constant c independent of K, where L is the total binary input length. The source width is O(k log n), so the reduction is fixed-parameter tractable in this standard input-length model. This does not preclude the pseudo-polynomial algorithm of Theorem 3, whose numerical range can grow as a power of n depending on K.

## 6. Fixed-r Factor Extension

Let r be fixed and 1≤K≤n. Each loading vector $\ell_i$ is in $\mathbb{Z}^r$. Let F be a symmetric positive-semidefinite rational r-by-r matrix and let $d_i$ be rational. All rationals are reduced and arithmetic is exact. Signed $d_i$ are allowed algebraically; the covariance interpretation requires $d_i\ge0$, and $d_i>0$ guarantees positive definiteness. Writing b(S) for the sum of the selected loading vectors,

$$F_r(S) = b(S)^{\mathrm{T}} F\, b(S) + \sum_{i\in S} d_i \qquad (7)$$

**Theorem 5 (fixed-r pseudo-polynomial frontier).** For fixed r, let $B_q=\sum_i|\ell_{iq}|$. There are $O(nK\prod_{q=1}^{r}(2B_q+1))$ transitions, each using poly(L) bit operations, where L is total input length. All stored rational values have bit lengths polynomial in L.

**Proof.** A state stores $(j,k,b_1,\ldots,b_r)$ and the minimum idiosyncratic sum. Coordinate q ranges from $-B_q$ to $B_q$. Include/exclude transitions are identical to Theorem 3; at the end evaluate $b^{\mathrm{T}}Fb$ plus the stored idiosyncratic term. For fixed r the number of loading-sum states is pseudo-polynomial in the magnitudes $B_q$. Sums and products of polynomially many rational inputs have polynomially bounded numerator/denominator bit growth, yielding the stated poly(L) factor. □

**Bit-complexity detail.** Write each input rational in lowest terms. A state value is a sum of at most n residual variances; before reduction, its denominator divides a product of at most n input denominators, whose bit length is bounded by the sum of their encoded bit lengths and hence by O(L). The associated numerator likewise has polynomial bit length. For each coordinate q, $B_q=\Sigma_i|\ell_{iq}|$ has O(L) bits. Because r is fixed, evaluating $b^{\mathrm{T}}Fb$ uses only a fixed number of products of O(L)-bit integers with input rationals and a fixed number of additions. Thus every stored state value and every final objective value has bit length polynomial in L.

For fixed r and polynomially bounded integer loadings, this is polynomial in total input length even when F and the residual variances are binary-encoded rationals. The claim does not extend to arbitrary rational loadings: clearing their denominators can make the integer state range exponentially large. Taking r=1 also gives a signed-loading dynamic program with loading sums from −B to B, where B is the sum of absolute loadings. Consequently, the integer D=I family in Theorem 6 is weakly NP-complete as well.

## 7. Signed-Loadings Structural Hardness Boundary with D = I

**Theorem 6 (signed-homoskedastic NP-completeness).** The exactly-K factor-encoded problem is NP-complete for $\Sigma=\beta\beta^{\mathrm{T}}+I$ even when exactly one loading is negative.

Proof. Reduce from positive Exact-k Subset Sum. Choose $H>\max_i a_i$. Create ordinary loadings $\beta_i=H+a_i$ and one anchor $\beta_0=-(kH+t)$. Set K=k+1, $d_i=1$ for all n+1 items, and decision threshold C=K. Then $F(S)=(\sum_{i\in S}\beta_i)^2+K$, so $F(S)\le K$ iff the selected loading sum is zero. Because all ordinary loadings are positive, every zero-sum solution includes $\beta_0$; exact cardinality leaves exactly k ordinary items, whose zero-sum condition is $\sum a_i=t$. Conversely every source solution plus $\beta_0$ is feasible. The construction has polynomial bit length and $\beta\beta^{\mathrm{T}}+I\succ0$. □

**Corollary 6.1 (sign-induced tractability-hardness boundary for D=I).** With D=I and $\beta_i \geq 0$, Theorem 1 gives O(n log n) comparisons; allowing a single negative loading already yields NP-completeness by Theorem 6.

## 8. Normalized Bit Projection and Restricted Circuit Lower Bounds

Circuit convention. $AC^0$ means nonuniform constant-depth circuits with unbounded-fan-in AND/OR gates and negations pushed to input literals. Circuit size counts non-input gates, matching Lin (2026).

Let $n \geq k \geq 2$ and $b \geq 1$ be integers, $Q=2^b$, and $z_1,\dots,z_n \in \{0,\dots,Q-1\}$. Modular k-SUM asks whether exactly k distinct indices have residue sum zero modulo Q. Repeated numerical values at distinct indices are allowed.

**Theorem 7 (normalized depth-zero projection).** For $n \geq k \geq 2$ and $b \geq 1$, modular k-SUM projects to the restriction of $L_1F$ with D=I and C=K. Every target coordinate is a source bit or a constant, and the output length depends only on (n,k,b).

Construction. Set $\beta_i = Q + z_i$ for each ordinary item and add anchors $\alpha_q = -(k+q)Q$ for $q=0,\dots,k-1$. Set $K'=k+1$, $d_i=1$ for all n+k items, and $C'=k+1$. The exact normalized loading width is $W\beta = 1 + \lceil \log_2((2k-1)Q) \rceil$; Wd=1 and WC=swidth(k+1). The largest-magnitude anchor attains the required width. Because $(2k-1)Q$ is not a power of two for $k \geq 2$, the negative endpoint does not create a one-bit discrepancy between signed positive and negative widths. All ordinary loadings lie in $[Q, 2Q-1]$ and fit within the same width. Their b low bits are the source bits and the remaining bits are constants. Every anchor bit and header bit is constant at fixed (n,k,b). Consequently every output coordinate copies one input bit or a constant, and every output is canonically normalized.

Correctness. Since D=I and $C'=K'$, acceptance is equivalent to selected loading sum zero. No solution uses zero anchors. If $r \geq 2$ anchors are chosen, the negative magnitude is at least rkQ, whereas fewer than $2(k+1-r)Q$ positive units remain, and

$$rkQ - 2(k+1-r)Q = [k(r-2) + 2(r-1)]Q > 0 \qquad (8)$$

so the total sum is negative. Hence exactly one anchor $\alpha_q$ and exactly k ordinary items are chosen. The zero-sum condition is

$$\sum_{i \in T} z_i = qQ \qquad (9)$$

which is equivalent to $\sum_{i \in T} z_i \equiv 0 \pmod{Q}$, because $0 \leq \sum z_i < kQ$ and $q \in \{0,\dots,k-1\}$. Conversely any modular solution determines q and gives a target zero-sum solution. □

### 8.1 Direct interface with the modular source

Lin (2026, Revision 2, Definition 3.1(iii), p. 18) defines k-SUM on indexed residues modulo $2^m$, with exactly k distinct indices and repeated values allowed. Thus its source is exactly the modular predicate in Theorem 7, with m=b. No representation conversion is required. Exact signed integer k-SUM, used separately in Theorem 4, must not be identified with this modular predicate without a no-wraparound argument.

For completeness, exact signed m-bit zero-sum instances also project to this source: sign-extend each two's-complement number to $b = m + \lceil \log_2 k \rceil + 2$ bits. Every k-term integer sum has magnitude less than $2^b$, so its residue is zero precisely when the integer sum is zero. This is a bit-copy map, but is not used in Corollary 7.1.

### 8.2 Target length and the fixed cardinality lower bound

Write $h(x) = 2\ell(x) + 1$ for the bit length of nat(x), where $\ell(x) = \max\{1, \lfloor \log_2 x \rfloor + 1\}$ for $x \geq 1$ and $\ell(0)=1$. The exact output length is $N = h(n+k) + h(k+1) + h(W\beta) + h(1) + h(WC) + (n+k)(W\beta+1) + WC$. In particular,

$$N = \Theta((n+k)(b + \log k) + \log n + \log b) \qquad (10)$$

Lin (2026, Theorem 5.1, equation (46), pp. 31–32) gives universal constants $\eta > 0$, $k_0$ and $C\Sigma$: for each fixed depth d and fixed $k \geq k_0$, the modular source needs at least $(n/k)^{\eta(k-1)}$ gates for all $n \geq n_1(d,k)$ and $b \geq C\Sigma k \log_2(en/k)$. The rate $\eta$ is independent of d and k. Choose $b = \lceil C\Sigma k \log_2(en/k) \rceil$, increasing $C\Sigma$ if necessary. This explicit choice, not the lower inequality alone, gives $N = \Theta(n \log n)$ for fixed k.

**Corollary 7.1 (transferred fixed-k lower bound).** With this choice of b, for every fixed d and $k \geq k_0$ and all sufficiently large n, every depth-d circuit correct on the Theorem 7 projection image requires at least $(n/k)^{\eta(k-1)}$ non-input gates. The same bound holds for a circuit deciding $L_1F$ on all N-bit strings. Proof. Substituting the output bits of Theorem 7 gives a circuit for Lin's modular predicate with no additional gates or depth. Apply the source theorem. □

**Corollary 7.2 ($AC^0$ nonmembership independent of Lin).** The language $L_1F$ is not in nonuniform $AC^0$. This already holds on its D=I restriction with polynomially bounded integer coefficients.

Proof. Given $t \geq 2$ bits $y_1, \ldots, y_t$, apply Theorem 7 with n=k=t, b=1 and $z_i = y_i$. The only t-subset of the source list is the whole list, so acceptance means that the number of 1 bits is even. The portfolio has 2t assets, K=t+1, ordinary loadings $2+y_i$, anchors $-2(t+q)$ for $0 \leq q < t$, residual variances 1, and threshold t+1. Its length is $N(t) = \Theta(t \log t)$; all coefficients have magnitude O(t). A polynomial-size constant-depth circuit family for these target lengths would therefore give polynomial-size constant-depth circuits for even parity, contradicting Furst, Saxe, and Sipser (1984, Theorem 3.3). Complementing the output interchanges even and odd parity and preserves $AC^0$. This argument uses no statement from Lin. □

The quantitative and qualitative conclusions concern different regimes. Corollary 7.1 fixes k before n grows. Corollary 7.2 takes k=n and uses the parity obstruction to constant depth, even though the numerical instances are solvable in polynomial time by the signed dynamic program. Alternatively, Corollary 7.1 implies full-language nonmembership by choosing fixed k with $\eta(k-1) > a+2$ against a hypothetical size bound $O(N^a)$. A polynomial-time algorithm and an $AC^0$ lower bound are fully compatible.

## 9. Unrestricted Circuit Profile $\Gamma_1$ F and the Standard P/poly Bridge

For unrestricted circuits, use fan-in at most two over {AND,OR,NOT}. Size counts AND/OR/NOT gates; input variables and the constants 0 and 1 are free. For every input length m define

$$\Gamma_1 F(m) = \min\{size(C) : C \text{ decides } L_1F \text{ on every } m\text{-bit string}\} \qquad (11)$$

**Proposition 8 (standard P/poly consequence).** $L_1F \in$ P/poly if and only if NP⊆P/poly.

**Proof.** If NP⊆P/poly, then $L_1F \in$ P/poly because $L_1F \in$ NP. Conversely, suppose $L_1F \in$ P/poly and let A∈ NP. Let R_A be a polynomial-time many-one reduction to $L_1F$ whose output length is at most q(n). For n-bit x, a polynomial-size circuit computes R_A(x) together with its output length ℓ. For every $\ell \leq q(n)$, use the polynomial-size $L_1F$ circuit C_ℓ; evaluate these polynomially many circuits in parallel on the appropriately wired output bits and select C_ℓ with a polynomial-size multiplexer controlled by the computed length. The total circuit size remains polynomial, so A∈ P/poly. Hence NP⊆P/poly. □

**Corollary 8.1 (exact interpretation of $\Gamma_1$F).** $\Gamma_1$F is not polynomially bounded iff NP⊄P/poly.

The phrase "not polynomially bounded" is used intentionally; it is the exact negation of membership in P/poly and avoids the stronger pointwise meaning sometimes attached to "super-polynomial."

**Corollary 8.2.** NP⊄P/poly implies P≠NP; the converse is not known.

## 10. Restricted Lower Bounds versus P/poly

Corollary 7.2 excludes nonuniform $AC^0$ using the classical parity theorem. Since $AC^0$ is a subclass of P/poly, this conclusion does not exclude P/poly. Proposition 8 and Corollary 8.1 characterize the unrestricted separation that remains unproved; this is a standard consequence of NP-completeness, not a new route around the known barriers to general circuit lower bounds.

## 11. Complexity-Theoretic Context and Possible Extensions

### 11.1 General lower-bound barriers

The restricted-circuit results above should be interpreted within the standard landscape of general circuit lower bounds. Relativization (Baker, Gill, & Solovay, 1975), the Natural Proofs framework (Razborov & Rudich, 1997), and algebrization (Aaronson & Wigderson, 2009) identify limitations of broad classes of proof techniques. They are not negative results about $L_1F$ itself and are not used in any proof in this paper; they serve only to explain why the $AC^0$ separation in Section 8 should not be extrapolated to P/poly.

### 11.2 Hardness magnification and meta-complexity

Hardness magnification for gap versions of MCSP and time-bounded Kolmogorov-complexity problems shows that modest lower bounds for selected meta-complexity problems can imply major separations (Oliveira, Pich, & Santhanam, 2021). Separately, Murray and Williams (2017) show that NP-hardness of MCSP under logtime-uniform $AC^0$ reductions would imply strong circuit lower bounds, including $NP \not\subset P/poly$. No reduction of either kind is established for $L_1F$ here. A meaningful extension would therefore require an explicit locality- and size-preserving connection to an appropriate meta-complexity problem rather than a generic appeal to hardness magnification.

## 12. Literature Positioning, Novelty, and Scope

### 12.1 Relation to cardinality-constrained and low-rank optimization

Bienstock (1996) and Bertsimas and Shioda (2009) develop algorithms for mixed-integer and cardinality-constrained quadratic optimization. Çela, Klinz, and Meyer (2006) explicitly note that constant-rank unconstrained quadratic 0–1 optimization is NP-hard already at rank one, using the classical SUBSET SUM square objective, and they give a pseudo-polynomial algorithm for constant rank. Accordingly, Theorem 2 is not presented as a first rank-one quadratic NP-hardness result, and Theorems 3 and 5 do not claim a new generic low-rank dynamic-programming principle. Theorem 2 instead establishes hardness under the simultaneous portfolio-specific restrictions of exact cardinality, factor-form input, positive integer loadings, strictly positive asset-specific residual variances, and positive definiteness, with direct containment in the EWCCMVSF formulation. Gao and Li (2013) identify a polynomial regime when all but a fixed number of the largest eigenvalues coincide. For $\beta\beta^T+I$ with nonzero $\beta$, the exceptional eigenvalue is the largest one, so that spectral condition does not automatically imply tractability of the signed family in Theorem 6.

Monge (2017, 2020) supplies the closest equally weighted factor-model formulation. Anis and Kwon (2025) consider decision-based learning for cardinality-constrained portfolio optimization. Ebrahimi, Amini, and Liu (2026, proof of Lemma 2.1) reduce CLIQUE using equal weights and a positive-semidefinite covariance obtained by shifting a graph matrix. That construction uses general covariance structure rather than the one-factor restrictions proved here. General portfolio hardness alone is therefore insufficient to establish Theorems 2 and 6.

### 12.2 Positioning of the present results

The paper's substantive claims are deliberately restricted to the stated model and encoding. Theorem 2 isolates a positive-integer, strictly positive-definite one-factor NP-complete family; Theorem 6 isolates identity residual covariance with exactly one negative loading; Theorem 4 preserves the natural cardinality parameter while reaching the positive-data family; and Theorem 7 gives an explicit normalized projection in which every output bit is a source bit or a constant. The fixed-k quantitative circuit consequence depends on Lin (2026), whereas the full-language $AC^0$ nonmembership follows independently from classical parity hardness.

The contribution is therefore the combination of portfolio-model restrictions, exact parameter preservation, and an encoding-local circuit interface. The K-smallest selection rule, SUBSET SUM completion of the square, sum-state dynamic programming, the parity lower bound, and the P/poly consequence of NP-completeness are established techniques. The paper specializes and composes these tools to obtain the stated one-factor classification while keeping the fixed-cardinality quantitative circuit statement separate from full-language $AC^0$ nonmembership and from the unresolved unrestricted-circuit frontier.

### 12.3 The Monge polynomial-solvability question

Monge (2017, Appendix, p. 22) explicitly asks whether the equality-weighted cardinality-constrained single-factor problem can be solved in polynomial time. Corollary 2.2 provides a complexity-theoretic resolution for the general binary-encoded variance-input exact-optimization formulation: a general exact polynomial-time algorithm would imply P=NP. Equivalently, under the standard assumption P≠NP, no such general exact polynomial-time algorithm exists. This conclusion leaves structured subclasses, including the nonnegative homoskedastic regime of Theorem 1, polynomially solvable.

### 12.4 Scope and limitations

This is a classification of specified regimes, not an exhaustive dichotomy for all sign patterns, rational loadings, side constraints, approximation criteria, or factor dimensions. Corollary 7.1 invokes an external 2026 ECCC theorem and inherits its stated width, depth, fixed-k, and onset conditions; Corollary 7.2 instead depends only on Theorem 7 and the classical parity lower bound. Neither circuit result excludes unrestricted polynomial-size circuits. The paper concerns exact worst-case computation and does not establish practical portfolio difficulty, forecasting performance, approximation hardness, average-case hardness, or a separation of P from NP.

## 13. Summary of Main Complexity Results

| **Regime / result** | **Assumptions** | **Complexity conclusion** |
|---|---|---|
| Nonnegative homoskedastic one-factor | $d_i=d>0$; $\beta_i\geq 0$; exact K | O(n log n) comparisons after reading the input (Theorem 1). |
| Positive asset-specific one-factor | $\beta_i,d_i\in\mathbb{Z}_+$; positive $d_i$ may vary by asset; $\Sigma\succ 0$ | NP-complete (Theorem 2). |
| Positive integer numerical regime | As in Theorem 3; $B=\Sigma_i\beta_i$ | Exact pseudo-polynomial DP: $O(nKB\cdot poly(L))$; weak numerical NP-completeness. |
| Cardinality parameter | Parameter K | W[1]-hard via a parameter-preserving reduction with K=k (Theorem 4). |
| Fixed factor dimension r | Integer loadings; fixed r; exact rational arithmetic | Pseudo-polynomial state-space algorithm $O(nK\prod q(2Bq+1)\cdot poly(L))$ (Theorem 5). |
| Signed homoskedastic one-factor | D=I; exactly one negative loading | NP-complete (Theorem 6). |
| Canonical factor language restricted to D=I | Normalized factor encoding; fixed-depth nonuniform $AC^0$ | Not in nonuniform $AC^0$ by a parity projection (Corollary 7.2); fixed-k bound via Lin (Corollary 7.1). |
| EWCCMVSF exact optimization | General rational/integer factor data | A general exact polynomial-time algorithm would imply P=NP (Corollary 2.2). |
| Unrestricted circuit profile $\Gamma_1 F$ | General Boolean circuits | $\Gamma_1 F$ is not polynomially bounded iff NP⊄P/poly; this frontier is open (Proposition 8/Corollary 8.1). |

## 14. Open Problems and Research Directions

Parameterized classification. Determine whether the W[1]-hardness result can be sharpened to W[1]-completeness for natural subfamilies, and derive ETH/SETH consequences or kernelization lower bounds under standard assumptions.

Factor-dimension dependence. Develop sharper algorithms in terms of r, the loading ranges, and the structure of the loading matrix. In particular, identify useful restrictions that reduce the product of coordinate ranges in Theorem 5. The fixed-r dynamic program alone gives no polynomial running-time guarantee when r grows.

Stronger restricted circuits. Investigate lower bounds beyond $AC^0$ while preserving the locality and size control of the reduction. Any such extension must keep the factor-input encoding explicit and must not be conflated with an unrestricted P/poly separation.

Meta-complexity connections. Determine whether $L_1F$ or a closely related canonical restriction admits a rigorously local reduction to or from Gap-MCSP, Gap-MKtP, or another hardness-magnification problem with sufficient parameter and size preservation to yield new circuit consequences.

## 15. Conclusion

Exact-cardinality, equally weighted one-factor selection exhibits a sharp contrast between transparent tractable structure and narrowly specified NP-complete families. Constant residual variance with nonnegative loadings permits selection of the K smallest loadings, while strictly positive asset-specific residual variances that vary by asset, or identity residual covariance with exactly one negative loading, suffice for worst-case NP-completeness. For Monge's 2017 variance-input formulation, the positive-data reduction shows that a general exact polynomial-time algorithm would imply P=NP.

The cardinality-preserving reduction establishes W[1]-hardness, while numerical dynamic programs describe one-factor and fixed-factor regimes. A normalized bit projection from modular k-SUM gives both a fixed-k quantitative circuit transfer and a separate parity-based proof of nonuniform $AC^0$ nonmembership. The latter uses only the classical parity lower bound and already holds on polynomially bounded numerical instances.

The results separate three issues that are often conflated: model structure, binary encoding, and circuit depth. They do not yield an unrestricted circuit lower bound. Proposition 8 and Corollary 8.1 merely record the standard fact that non-polynomial boundedness of the unrestricted circuit profile $\Gamma_1F$ is equivalent, for this NP-complete language, to $NP \not\subset P/poly$. Thus the paper's unconditional circuit contribution remains precisely the restricted-model separation proved in Section 8.

## Appendix A. Reduction Sizes and Encoding Bounds

| Reduction | Target items | Parameter | Magnitude/width control | Time | Encoding property |
|---|---|---|---|---|---|
| Lemma 2.1 | 2n | K=n | Shifted $a_i+1$; target t+n | Polynomial | Canonical integers |
| Theorem 2 | n | K=k | $M=2ta_{max}+t^2+1$ | Polynomial | Factor input |
| Theorem 4 | n | K=k | $\lvert x_i\rvert \le n^{2k}$; O(k log n) source bits | FPT/poly | Parameter preserving |
| Theorem 6 | n+1 | K=k+1 | one negative anchor; D=I | Polynomial | Factor input |
| Theorem 7 | n+k | K′=k+1 | normalized Wβ=f(k,b); Wd=1; WC=f(k) | Projection | Every target bit source/constant |

## Appendix B. Canonical Encoding Grammar

Header = nat(n)||nat(K)||nat(Wβ)||nat(Wd)||nat(WC). Data = $\beta_1$||…||$\beta_n$||$d_1$||…||$d_n$||C. The β-blocks are two's-complement signed Wβ-bit strings; d-blocks are Wd-bit unsigned strings; C is WC-bit two's complement. Wβ, Wd,

WC must equal the minimal normalized widths defined in Section 1.1. Any alternative padding is malformed and rejected. This normalization prevents arbitrary length inflation of the same numerical instance and makes $\Gamma_1 F$ length-sensitive in a fixed, reproducible way.

## Declarations

Author contributions. Davit Gondauri is the sole author and is responsible for the conceptualization, formal analysis, methodology, writing, and revision of the manuscript.

Data availability. No empirical dataset is used in this theoretical study.

Code availability. The results are established by analytical proofs; no implementation or computational dataset is required to apply the stated reductions and recurrences.

**Ethics statement.** Not applicable. This is a theoretical study and uses no human participants, animals, personal data, or empirical intervention.

## References

Abboud, A., Lewi, K., & Williams, R. (2014). Losing Weight by Gaining Edges. In Algorithms – ESA 2014 (LNCS 8737, pp. 1–12). Springer. https://doi.org/10.1007/978-3-662-44777-2_1

Aaronson, S., & Wigderson, A. (2009). Algebrization: A new barrier in complexity theory. ACM Transactions on Computation Theory, 1(1), Article 2. https://doi.org/10.1145/1490270.1490272

Anis, H. T., & Kwon, R. H. (2025). End-to-end, decision-based, cardinality-constrained portfolio optimization. European Journal of Operational Research, 320(3), 739–753. https://doi.org/10.1016/j.ejor.2024.08.030

Baker, T., Gill, J., & Solovay, R. (1975). Relativizations of the P =? NP question. SIAM Journal on Computing, 4(4), 431–442. https://doi.org/10.1137/0204037

Bertsimas, D., & Shioda, R. (2009). Algorithm for cardinality-constrained quadratic optimization. Computational Optimization and Applications, 43(1), 1–22. https://doi.org/10.1007/s10589-007-9126-9

Bienstock, D. (1996). Computational study of a family of mixed-integer quadratic programming problems. Mathematical Programming, 74(2), 121–140. https://doi.org/10.1007/BF02592208

Çela, E., Klinz, B., & Meyer, C. (2006). Polynomially solvable cases of the constant rank unconstrained quadratic 0–1 programming problem. Journal of Combinatorial Optimization, 12(3), 187–215. https://doi.org/10.1007/s10878-006-9625-0

Ebrahimi, A., Amini, H., & Liu, H. (2026). Cardinality-constrained portfolio optimization with clustering. Annals of Operations Research. https://doi.org/10.1007/s10479-026-07184-z

Furst, M., Saxe, J. B., & Sipser, M. (1984). Parity, circuits, and the polynomial-time hierarchy. Mathematical Systems Theory, 17, 13–27. https://doi.org/10.1007/BF01744431

Gao, J., & Li, D. (2013). A polynomial case of the cardinality-constrained quadratic optimization problem. Journal of Global Optimization, 56(4), 1441–1455. https://doi.org/10.1007/s10898-012-9853-z

Garey, M. R., & Johnson, D. S. (1979). Computers and Intractability: A Guide to the Theory of NP-Completeness. W. H. Freeman.

Lin, H. (2026). Fine-Grained $AC^0$ Lower Bounds for k-OV, k-XOR, and k-SUM via Colored Subgraph Isomorphism. ECCC TR26-139, Revision 2, 4 September 2026. https://eccc.weizmann.ac.il/report/2026/139/

Monge, J. F. (2017). Cardinality constrained portfolio selection via factor models. arXiv:1708.02424v1. https://arxiv.org/abs/1708.02424v1

Monge, J. F. (2020). Equally weighted cardinality constrained portfolio selection via factor models. Optimization Letters, 14, 2515–2538. https://doi.org/10.1007/s11590-020-01571-6

Murray, C. D., & Williams, R. R. (2017). On the (Non) NP-Hardness of Computing Circuit Complexity. Theory of Computing, 13(4), 1–22. https://doi.org/10.4086/toc.2017.v013a004

Oliveira, I. C., Pich, J., & Santhanam, R. (2021). Hardness Magnification Near State-of-the-Art Lower Bounds. Theory of Computing, 17(11), 1–38. https://doi.org/10.4086/toc.2021.v017a011

Razborov, A. A., & Rudich, S. (1997). Natural Proofs. Journal of Computer and System Sciences, 55(1), 24–35. https://doi.org/10.1006/jcss.1997.1494